\documentclass[twocolumn,preprintnumbers,amsmath,amssymb]{revtex4}

\usepackage{graphicx}
\usepackage{dcolumn}
\usepackage{bm}
\begin{document}


\title{Quantum Wave Mechanics as the Magnetic Interaction of Dirac Particles}



\author{David C. Lush}


\affiliation{%
 d.lush@comcast.net \\}%


\date{\today}

\begin{abstract}

It is shown that a wave mechanical quantum theory can be derived from relativistic classical electrodynamics, as a feature of the magnetic interaction of Dirac particles modeled as relativistically circulating point charges.  The magnetic force between two classical point charges, each undergoing relativistic circulatory motion of small radius compared to the separation between their centers of circulation, and assuming a time-symmetric electromagnetic interaction, is modulated by a factor that behaves similarly to the Schr\"odinger wavefunction. The magnetic force between relativistically-circulating charges has been shown previously to have a radially-directed inverse-square part of similar strength to the Coulomb force, and sinusoidally modulated by the phase difference of the charges' circulatory motions. The magnetic force modulation in the case of relatively moving centers of charge circulation solves an equation formally identical to the time-dependent free-particle Schr\"odinger equation, apart from a factor of two on the partial time derivative term. Considering motion in a time-independent potential obtains that the modulation also satisfies an equation formally similar to the time-independent Schro\"dinger equation. Using a formula for relativistic rest energy advanced by Osiak, the time-independent Schr\"odinger equation is solved exactly by the resulting modulation function.  The significance of the quantum mechanical wavefunction follows straightforwardly from these observations.  After considering the modification of Wheeler-Feynman absorber theory required by the adoption of Minkowski-Osiak relativity, the model is extended to obtain the full complex Schr\"odinger wavefunction.

\end{abstract}

\maketitle

\begin{widetext}
\tableofcontents
\end{widetext}

\pagebreak

\section{Introduction}
\label{intro}


When de Broglie hypothesized a wave character of matter \cite{deBroglieCRN,deBroglieDT}, he supposed elementary particles of matter have an internal oscillation of fixed frequency in the particle rest frame, equal that of a photon of the matter particle's relativistic energy. By equating the phase history of an elementary particle internal oscillation to a superluminal phase velocity of a conjectural ``phase wave'' traveling with group velocity equal the particle velocity, de Broglie proceeded to derive a travelling wave with photon-like wavelength and frequency-to-energy relationships. The physical importance of de Broglie's matter-associated wave was subsequently made certain when Schr\"odinger showed \cite{Schrodinger1926c} a complex generalization of it satisfies a wave equation that predicts the emission spectrum of the hydrogen atom with unprecedented precision. 


More recently, various authors \cite{Huang1952,BarutBracken,BarutZhangi} have noted that the {\em zitterbewegung}  \cite{Schrodinger1930} of the Dirac electron theory \cite{Dirac1928} can be related to the  spin of the electron, and provisionally \cite{Hestenes1990} to the internal oscillation envisioned by de Broglie. It's possible to interpret the zitterbewegung, or ``jittery motion,'' of the Dirac electron as a circulatory relativistic motion of a point charge, such that the charge orbital angular momentum around its average position constitutes the electron intrinsic spin \cite{SalesiRecamiY10}.  This is the ``zitter'' \cite{HestenesY10} model of the electron.  Classical electrodynamics in the highly relativistic limit, which, strictly, requires the use of functional rather than ordinary differential equations for proper description \cite{RajuY5}, has been shown to lead to such motions \cite{DeLucaY6}, and internal particle motions  or the zitterbewegung can be linked to both the de Broglie-Bohm \cite{HollandY14,HileyY14,SalesiY9} and Elementary Cycles \cite{DolceY12,DolceY16} interpretations of quantum theory.

In  an earlier paper \cite{LushY16}, using prior results by Rivas \cite{Rivas,RivasY7}, it was described how magnetic interactions between zitter particles, {\emph i.e}, elementary particles consisting of relativistically circulating classical point charges, can have a Coulomb-like radial inverse-square force component, having similar strength to the Coulomb force but with modulation that can be related to the de Broglie wavelength.  Interpretation of the phase of the de Broglie matter wave or Schr\"odinger wavefunction as the difference, observed via electromagnetic interaction, of the zitterbewegung phases of two interacting particles, rather than as simply an internal oscillation of an individual particle, provides a plausible classical physics explanation for quantum behavior.  As will be shown, the modulation of the force between two zitter particles interacting magnetically and time-symmetrically satisfies the Schr\"odinger equation if \(\hbar\) is replaced by \(\hbar/2\). If the hypothesis advanced by Osiak is correct that the mass-energy equivalency is properly \(E = m c^2/2\), then the modulation solves the Schr\"odinger equation exactly.   

\section{Zitter Particle Model and Inverse-Square Radial Magnetic Force Between Zitter Particles}
\label{sect_ret_smallv}

The zitter particle model, proposed by Hestenes \cite{Hestenes1990}, represents intrinsic spin as orbital angular momentum of a point-like particle undergoing circulatory motion at speed \(c\).  An explanation for the existence of such circularatory motion is outside the scope of the zitter model.  However, such circulatory motion can arise naturally if the electron and other elementary particles are bound states of smaller particles \cite{LincolnY12} as proposed independently by Shupe \cite{Shupe1979} and Harari \cite{Harari1979}.

\subsection{De Broglie Matter Wave and Zitter Model of Dirac Particle}
\label{ZitterModel}

De Broglie's hypothesis \cite{deBroglieCRN} of a wave character of matter was based on analogy with the wave character of light.  He proposed an internal oscillatory property of matter, that created a wave phase due to  motion of an elementary matter particle, by combining the Einstein mass-energy equivalency \(E = m c^2 = \gamma m_0 c^2\), where \(\gamma\) is the Lorentz factor, with the Planck-Einstein law \(E = h \nu\).  De Broglie proposed an internal frequency could be associated with an elementary particle based on its rest mass \(m_0\) as
\begin{equation}
\nu_B  = \frac{E_0}{h} = \frac{m_0 c^2}{h} \equiv \frac{\omega_B}{2 \pi}, 
\label{debroglieclock}
\end{equation}  
where \(h\) is Planck's constant, and \(c\) is the ``limiting velocity of the theory of relativity''.  The hypothesis is difficult to reconcile with the frequency-scaling behavior of light and photons, however. While for a photon, frequency is proportional to energy, any internal motion of a moving massive particle is perceived by a stationary observer to be slowed by time dilation.

As de Broglie showed \cite{deBroglieDT}, it's nonetheless possible to derive a modulation with the necessary frequency scaling behavior from the hypothesis of Eq. (\ref{debroglieclock}), by equating the phase history generated by fixing the phase angle of a sinusoidal modulation with frequency \(\nu_B\) to the time-varying position of the particle with the phase of a traveling wave of superluminal phase velocity. The group velocity of the resulting wave is equal to the particle velocity, and the resulting wavelength depending inversely on the particle momentum is the quantity now called the de Broglie wavelength.  The de Broglie wavelength formula can be found alternatively, however, without reference to an internal motion, simply by combining the Planck-Einstein law with the relativistic description of momentum.  In either case, the Planck-Einstein relation is taken as a postulate.  

The de Broglie matter wave, or ``phase wave,'' can be represented in terms of a spatial coordinate \(x\) and time \(t\) as \cite{deBroglieDT}
\begin{equation}
B(x,t) = \cos(kx - \omega t)  = \cos \gamma\omega_B(x/\upsilon_B -  t),
\end{equation}
with angular frequency \(\omega = \gamma\omega_B = 2\pi\gamma\nu_B\), where \(\nu_B\) is the de Broglie frequency as given by Eq. (\ref{debroglieclock}), phase velocity \(\upsilon_B = \omega/k = c/\beta = c^2/\upsilon\), where \(\upsilon = c\beta\) is the velocity of the particle associated with the phase wave, and the Lorentz factor \(\gamma = (1 - \beta^2)^{-1/2}\).  The wavenumber is thus \(k = 2 \pi \gamma \nu_B/\upsilon_B = 2 \pi \gamma \nu_B \beta / c = 2 \pi \gamma  m_0 c^2 \beta /(h c) = 2 \pi \gamma  m_0 \upsilon / h = 2 \pi p / h\), and the wavelength is \(\lambda = 2\pi/k = h/p\). As de Broglie shows, the group velocity can be calculated as \(\upsilon_g = d\omega/dk\) to obtain that \(\upsilon_g = \upsilon\), the particle velocity.

In the zitter particle model, the intrinsic angular momentum of a spin-half particle is modeled as orbital angular momentum of a classical point charge undergoing luminal-velocity circular motion around a fixed point in the particle rest frame.  The zitter particle model thus represents the particle position as an average of the charge position over a cycle of the circulatory motion, henceforth the zitter motion. The circular frequency of the zitter motion will be referred to as the zitter frequency. With angular momentum of a charge with rest mass \(m\) due to the zitter motion of radius \(r_0\) given as \(L = r_0 \gamma m \upsilon \) with \(\upsilon\) the velocity of the circulatory motion, and taking \(\gamma m = m_0\) where \(m_0\) is the observed particle rest mass, then in the limit of \(\upsilon=c\), zitter motion radius \(r_0  = \hbar / 2 m_0 c\) obtains that the particle intrinsic angular momentum is \(\hbar/2\). The diameter of the circulatory motion that has angular momentum \(\hbar/2\) is thus one reduced Compton wavelength, \(\lambda_0 = \hbar / m_0 c\). Denoting the zitter frequency of a spin-half particle by \(\nu_z\), then
\begin{equation}
\nu_z \equiv \frac{\omega_z}{2\pi}  = \frac{c}{2 \pi r_0} =  \frac{c}{\pi \lambda_0} = \frac{2 m_0 c^2}{h} = 2 \nu_B,
\label{zitter_freq_spin_half}
\end{equation}
where \(\nu_B\) is the de Broglie frequency as given by Eq. (\ref{debroglieclock}).  The frequency \(\nu_z\) is also the frequency of the zitterbewegung of the Dirac electron.

\subsection{Coulomb-Like Magnetic Force Between Dirac Particles}
\label{CoulombLikeForce}

In \cite{LushY16} it is shown that the magnetic force between zitter particles will have a radially-directed magnetic force component given by 
\begin{equation}
\mbox{\boldmath$F$}_r =  -\frac{q_t q_s \hat{\mbox{\boldmath$r$}}}{r^2} \mbox{\boldmath$\beta$}_t \cdot\left[\frac{\mbox{\boldmath$\beta$}_s}{\left(1 - \mbox{\boldmath$\beta$} \cdot \mbox{\boldmath$n$}\right)^3 (1 - 2\epsilon\hat{\mbox{\boldmath$r$}} \cdot \hat{\mbox{\boldmath$k$}} + {\epsilon}^2 )^{3/2}}\right]_{\textnormal{{\scriptsize ret}}}, 
\label{MagForceRadialTerm}
\end{equation}
where \(\mbox{\boldmath$\beta$}_s = \mbox{\boldmath$\upsilon_s$}/c\), with \(\mbox{\boldmath$\upsilon_s$}\) the velocity of a field-source zitter particle in circular motion around a stationary position at or near the speed of light. The motion of the second zitter particle referred to here as the test zitter particle is represented by  \(\mbox{\boldmath$\beta$}_t = \mbox{\boldmath$\upsilon_t$}/c\). The separation between the source zitter particle center of circulatory motion and the test zitter particle charge position is \(r\), with \(\hat{\mbox{\boldmath$r$}}\) a unit vector directed toward the test zitter particle charge position from the field-source zitter particle center of motion, and \(\epsilon = \lambda_0/r\) with \(\lambda_0\) the reduced Compton wavelength. \(\hat{\mbox{\boldmath$k$}}\) is a unit vector toward the source zitter particle charge position from the center of its circular motion. The test and source particle charges are \(q_t\) and \(q_s\).  The subscript ``ret'' indicates that quantities in the brackets are evaluated at the retarded time. (Strictly, \(r\) here is the distance from the source zitter particle center of motion to test particle charge position, but under the assumption of zitter particles separation large compared to the zitter motion radius, it may be taken as the separation between zitter centers.) For large enough interparticle separation, \(\epsilon << 1\) and so
\begin{equation}
\mbox{\boldmath$F$}_r \approx  \frac{-q_t q_s \mbox{\boldmath$\beta$}_t \cdot   \mbox{\boldmath$\beta$}_{s,{\textnormal{{\scriptsize ret}}}} \hat{\mbox{\boldmath$r$}}}{\left(1 - \mbox{\boldmath$\beta$}_{s,{\textnormal{{\scriptsize ret}}}} \cdot \mbox{\boldmath$n$}\right)^3 r^2},
\label{radial_force_term}
\end{equation}
where \(\mbox{\boldmath$\beta$}_s\) is rewritten as \(\mbox{\boldmath$\beta$}_{s,{\textnormal{{\scriptsize ret}}}}\) to emphasize its association with the source particle and at the retarded time.  The term given by Eq. (\ref{MagForceRadialTerm}) and for large enough interparticle separation by Eq. (\ref{radial_force_term}), that contributes to the magnetic force caused by a stationary zitter particle on another zitter particle, differs from the Coulomb force caused by a stationary charge on another classical point charge by the factor \(\mbox{\boldmath$\beta$}_t \cdot \mbox{\boldmath$\beta$}_{s,{\textnormal{{\scriptsize ret}}}}\)  and the inverse factor of \((1 - \mbox{\boldmath$\beta$}_{s,{\textnormal{{\scriptsize ret}}}} \cdot \mbox{\boldmath$n$})^3\).  The inverse factor of \((1 - \mbox{\boldmath$\beta$}_{s,{\textnormal{{\scriptsize ret}}}} \cdot \mbox{\boldmath$n$})^3\)  modulates the magnitude of the radial magnetic force at the frequency of the zitter motion, but averages to unity over a zitter cycle. For the zitter particles with luminal motion of the internal charges the magnitudes of \(\mbox{\boldmath$\beta$}_t\) and \(\mbox{\boldmath$\beta$}_s\) are unity, and for aligned in-phase zitter motions \(\mbox{\boldmath$\beta$}_t \cdot \mbox{\boldmath$\beta$}_s\) is also unity. Therefore \(\mbox{\boldmath$F$}_r\) where it is defined is a radially-directed inverse-square force with average strength equal to the Coulomb force between static charges, but modulated by the phase difference between the zitter motions, including phase delay due to propagation. It is thus justified to refer to the force as described by Eqs. (\ref{MagForceRadialTerm}) and (\ref{radial_force_term}) as the Coulomb-like magnetic force.

\section{Relative Motion Induced Modulation of the Coulomb-Like Magnetic Force}
\label{sect_ret_smallv}

In \cite{LushY16}, it is shown that the spatial period of modulation of the inverse-square radial magnetic force between zitter particles varies with particles' relative momentum in a manner that can be related to the de Broglie wavelength.  However, the correspondence in the field-source zitter particle rest frame as described in \cite{LushY16} is valid only in the limit of small test zitter particle velocity. The correspondence was limited, in part, due to omission of the full relativistic description of the field-source particle zitter phase as it influences the motion of the test zitter particle. The simple description based on time dilation used in \cite{LushY16} is insufficient.  Rather, a proper description must be based on Lorentz transformation of the source particle zitter phase from the laboratory frame to the test particle rest frame. As noted by Baylis \cite{BaylisY7}, the de Broglie matter or phase wave arises naturally by Lorentz transformation of a phase function representing a distributed synchronized system of laboratory frame clocks from the laboratory frame to the rest frame of a moving particle. This situation corresponds in the interacting zitter particles model to nonradial relative motion, where the relativistic Doppler shift is equivalent to time dilation. The interacting zitter particles model however involves the propagation delay and advancement from the field-source to the test particle, and has an electrodynamical basis and effect.   

In this section a detailed description is developed of the magnetic interaction of two equal-mass zitter particles with aligned spins, for the case of a stationary field-source spin-half zitter particle. The description is developed initially considering only retarded interaction, then the advanced interaction is included as needed for a time-symmetric  interaction. In the section following, the precise relationship between the de Broglie wavelength and the spatial modulation of the Coulomb-like magnetic force is developed.   In the limit of large velocity, the modulation spatial period approaches  one-half the de Broglie wavelength \cite{LushY19a} there due to the zitterbewegung frequency of the spin-half Dirac particle being twice the frequency de Broglie hypothesized based on the spin-one photon. In spite of being based on the photon, de Broglie's wavelength value has been demonstrated to be applicable to electron wavefunctions through the success of nonrelativistic wave mechanics. Therefore it is of interest that the nonrelativistic case obtains a modulation spatial period that is equal to the de Broglie wavelength for both nonradial and radial relative motion, assuming a time-symmetric electromagnetic field.

\subsection{Motion Induced Modulation for Retarded Interaction}
\label{ModulationRetarded}

Suppose that the test zitter particle is moving with velocity
\begin{equation}
\upsilon = \| \mbox{\boldmath$\upsilon$} \| = c \| <\mbox{\boldmath$\beta$}_t >\|,
\end{equation}
where the brackets indicate an average over a test particle zitter period, in the inertial reference frame where the  field-source zitter particle is stationary. Then, the phase of the test particle zitter motion as observed in the source zitter particle rest frame can be written as \cite{BaylisY7} \(\omega_z \tau = \gamma \omega_z ( t - \mbox{\boldmath$\upsilon$} \cdot \mbox{\boldmath$r$}/c^2)\), where \(\tau\) is the time coordinate in the test zitter particle rest frame, \(\gamma = (1 - (v/c)^2)^{-1/2}\), and \(\mbox{\boldmath$r$}\) here is the displacement from the source zitter particle to the test zitter particle. (In \cite{BaylisY7}, the frequency referred to is \(\omega_0 = m_0 c^2 / \hbar \equiv 2\pi \nu_B\) here rather than \(\omega_z \equiv 2\pi \nu_z = 2 m_0 c^2 / \hbar \), but the equation is equally applicable using the zitter frequency \(\omega_z\).) Representing the source particle zitter motion velocity in the x-y plane of its rest frame as \(\mbox{\boldmath$\beta$}_s = [-\sin(\omega_z t),\cos(\omega_z t),0]\), and assuming aligned spins and equal masses of the source and test zitter particles, the modulating factor on the inverse-square radial magnetic force in the numerator of the right-hand side of Eq. (\ref{radial_force_term}) becomes \footnotemark[1]
\footnotetext[1]{The modulating factor on the inverse-square part of the radially-directed magnetic force acting on the test zitter particle, for the case of parallel spins as modeled by the zitter motions, is given by Eq. (21) of \cite{LushY16}, incorrectly, as
\begin{equation}
 \mbox{\boldmath$\beta$}_t \cdot   \mbox{\boldmath$\beta$}_{s,{\textnormal{{\scriptsize ret}}}} = \cos(\omega' t - \omega t_{\textnormal{{\scriptsize ret}}} - \phi_0), 
\nonumber 
\end{equation}
where the zitter frequency of the test particle as observed in the source zitter particle rest frame was  stated as \(\omega' = \omega/\gamma \), with \(\omega = 2 m_e c^2 / \hbar\). This form accounts for time dilation but does not include the modulation phase variation with velocity and position as described by Eq. (\ref{retarded_modfact_tret}) above.}

\begin{eqnarray}
 \mbox{\boldmath$\beta$}_t \cdot   \mbox{\boldmath$\beta$}_{s,{\textnormal{{\scriptsize ret}}}} \nonumber \\ =   [-\sin\gamma \omega_z ( t - \mbox{\boldmath$\upsilon$} \cdot \mbox{\boldmath$r$}/c^2) ,\cos \gamma \omega_z ( t - \mbox{\boldmath$\upsilon$} \cdot \mbox{\boldmath$r$}/c^2) ,0] \nonumber \\ \cdot [-\sin(\omega_z t_{\textnormal{{\scriptsize ret}}} + \phi_0),\cos(\omega_z t_{\textnormal{{\scriptsize ret}}} + \phi_0),0],  
\label{first_correction} 
\end{eqnarray}
where \(\phi_0\) is a constant to account for the phase difference between the source and test particle zitter motions at (source particle rest frame) time \(t=0\).  Carrying out the multiplication and applying the trigonometric identity \(\sin \alpha \sin \beta + \cos \alpha \cos \beta = \cos (\alpha - \beta)\) obtains
\begin{equation}
 \mbox{\boldmath$\beta$}_t \cdot   \mbox{\boldmath$\beta$}_{s,{\textnormal{{\scriptsize ret}}}} =  \cos \left[\gamma \omega_z( t - \mbox{\boldmath$\upsilon$} \cdot \mbox{\boldmath$r$}/c^2) - \omega_z t_{\textnormal{{\scriptsize ret}}} - \phi_0\right]. 
\label{retarded_modfact_tret}
\end{equation}

In what follows the abbreviation \( {\cal M}(t) \equiv  \mbox{\boldmath$\beta$}_t(t) \cdot \mbox{\boldmath$\beta$}_s(t)\), where \(t\) is the time coordinate in the source zitter particle rest frame, will be used for the modulating factor generally.  The modulating factor on the Coulomb-like magnetic force acting on the test zitter particle due to the retarded magnetic field is then \( \mbox{\boldmath$\beta$}_t(t) \cdot \mbox{\boldmath$\beta$}_s ({t_{\textnormal{{\scriptsize ret}}}}) \equiv  \mbox{\boldmath$\beta$}_t \cdot   \mbox{\boldmath$\beta$}_{s,{\textnormal{{\scriptsize ret}}}}(t) \equiv {\cal M}_{{\textnormal{{\scriptsize ret}}}}(t)\).

Substituting for \(t_{\textnormal{{\scriptsize ret}}} =  t - r_{\textnormal{{\scriptsize ret}}}/c \) in Eq. (\ref{retarded_modfact_tret}),
\begin{eqnarray}
{\cal M}_{{\textnormal{{\scriptsize ret}}}}(t) =  \cos \left[\gamma \omega_z ( t - \mbox{\boldmath$\upsilon$} \cdot \mbox{\boldmath$r$}/c^2) - \omega_z (t - r_{\textnormal{{\scriptsize ret}}}/c) - \phi_0\right] \nonumber \\
 =  \cos \left[ (\gamma - 1) \omega_z t - \gamma \omega_z \mbox{\boldmath$\upsilon$} \cdot \mbox{\boldmath$r$}/c^2 + \omega_z (r_{\textnormal{{\scriptsize ret}}}/c) - \phi_0 \right]. 
\nonumber
\end{eqnarray}

Since the source zitter particle is assumed stationary here, \(r_{\textnormal{{\scriptsize ret}}}(t) = r(t) \equiv r\), and so (using the present notation for the zitter frequency generally as \(\omega_z\)),
\begin{equation}
{\cal M}_{{\textnormal{{\scriptsize ret}}}}(t) = \cos ( \omega_z \left[ (\gamma - 1) t - \gamma  \mbox{\boldmath$\upsilon$} \cdot \mbox{\boldmath$r$}/c^2 +  r/c\right] - \phi_0) , 
\label{modfact_retarded} 
\end{equation}
where  \(t\) is time in the source zitter particle rest frame. Here \(\gamma = (1 - \beta^2)^{-1/2}\), with \(\upsilon = c\beta = c\|<{\mbox{\boldmath$\beta$}_t}>\|\)  the test zitter particle velocity magnitude, the average over its charge zitter period.

The case of the test and source zitter particles having opposing spins is addressed by reversing the direction of the zitter motion of the test particle so that the modulation factor as given previously for aligned spins by Eq. (\ref{first_correction}) becomes, with  \(\mbox{\boldmath$\beta$}'_t \equiv  [\sin\gamma \omega_z( t - \mbox{\boldmath$\upsilon$} \cdot \mbox{\boldmath$r$}/c^2) ,\cos \gamma \omega_z( t - \mbox{\boldmath$\upsilon$} \cdot \mbox{\boldmath$r$}/c^2) ,0]\) representing the oppositely-spinning test particle,
\begin{eqnarray}
 \mbox{\boldmath$\beta$}'_t \cdot   \mbox{\boldmath$\beta$}_{s,{\textnormal{{\scriptsize ret}}}} \nonumber \\ =  [\sin\gamma \omega_z( t - \mbox{\boldmath$\upsilon$} \cdot \mbox{\boldmath$r$}/c^2) ,\cos \gamma \omega_z( t - \mbox{\boldmath$\upsilon$} \cdot \mbox{\boldmath$r$}/c^2) ,0] \nonumber \\ \cdot [-\sin(\omega_z t_{\textnormal{{\scriptsize ret}}} + \phi_0),\cos(\omega_z t_{\textnormal{{\scriptsize ret}}} + \phi_0),0]. 
\label{opposing_spins1} 
\end{eqnarray}

Applying the trigonometric identity \(-\sin \alpha \sin \beta + \cos \alpha \cos \beta = \cos (\alpha + \beta)\) obtains
\begin{equation}
 \mbox{\boldmath$\beta$}_t \cdot   \mbox{\boldmath$\beta$}_{s,{\textnormal{{\scriptsize ret}}}} =  \cos \left[\gamma \omega_z ( t - \mbox{\boldmath$\upsilon$} \cdot \mbox{\boldmath$r$}/c^2) + \omega_z t_{\textnormal{{\scriptsize ret}}} + \phi_0\right]. 
\label{retarded_modfact_tret_opp}
\end{equation}

Substituting for \(t_{\textnormal{{\scriptsize ret}}} =  t - r_{\textnormal{{\scriptsize ret}}}/c \) in Eq. (\ref{retarded_modfact_tret}), and for the stationary source zitter particle so that \(r_{\textnormal{{\scriptsize ret}}} =  r \),
\begin{eqnarray}
 \mbox{\boldmath$\beta$}_t \cdot   \mbox{\boldmath$\beta$}_{s,{\textnormal{{\scriptsize ret}}}} \equiv  {\cal M}_{{\textnormal{{\scriptsize ret}}}}(t)  \nonumber \\ =  \cos \left[\gamma \omega_z ( t - \mbox{\boldmath$\upsilon$} \cdot \mbox{\boldmath$r$}/c^2) + \omega_z (t - r_{\textnormal{{\scriptsize ret}}}/c) + \phi_0\right] \nonumber \\
  = \cos ( \omega_z \left[ (\gamma + 1) t - \gamma  \mbox{\boldmath$\upsilon$} \cdot \mbox{\boldmath$r$}/c^2 -  r/c\right] + \phi_0). 
\label{modfact_retarded_opposite} 
\end{eqnarray}

Comparing the modulation factor for opposing spins of Eq. (\ref{modfact_retarded_opposite}) with that of Eq. (\ref{modfact_retarded}) for aligned spins, the most significant difference apparent is that the frequency quantity \(\omega_z(\gamma - 1)\) for aligned spins becomes \(\omega_z(\gamma + 1)\) for opposing spins.  This is illustrative of the nature of the current model, that in addition to being interesting only for equal mass particles, it also must be restricted to aligned spins. For aligned spins and equal mass particles, the quantity \(\omega_z(\gamma - 1)\) can be associated directly with the test zitter particle relativistic kinetic energy. The relativistic kinetic energy of a particle of invariant mass \(m_0\) may be stated as \(K = E - E_0 = \gamma m_0 c^2 - m_0 c^2 = m_0 c^2 (\gamma - 1) = (\hbar/2)\omega_z (\gamma - 1)\). Thus, in the case of equal-mass zitter particles of aligned spins, the partial time derivative of the magnetic force on a moving zitter particle by a stationary one can be directly related to the kinetic energy of the moving particle.  This observation generalizes straightforwardly in the case of two moving particles to the difference of the relativistic energies of the two magnetically-interacting zitter particles.  On the other hand, for opposing spins, the frequency of the modulation is proportional to the sum of the total relativistic energies of the two particles, which cannot convey significant information about the dynamics in the nonrelativistic limit (or generally). 

A problem for the present model, then, is how to recover similar information about the interparticle dynamics in the case of opposing spins, to that available for aligned spins. One possibility is if particles such as electrons and quarks, usually regarded as elementary, are actually composites of more-fundamental particles. According to the preon model of elementary particles \cite{Shupe1979,Harari1979}, electrons are made up of three spin-half preons each with one-third the electron charge \cite{LincolnY12}. An electron composed of three spin-half constituents provides a possible mechanism for atomic-scale magnetic binding in the case of opposite spins.  While the radial inverse square law magnetic force has a vanishing average for opposite spins of interacting zitter particles consisting of single charges, the average magnetic interaction between electrons consisting of three spin-half constituents need not vanish for opposite net spins. Nonetheless, it seems unlikely that an interaction of composite electrons each consisting of three spin-half preons is consistent with observation, or that flipping an electron spin results in only a small correction to the nominal binding force.  For example, in the case of opposing net spins, the test electron will move coherently in the magnetic field of the two preons moving with equal orbital angular momentum around the (assumed stationary here) field-source electron center of mass. It is natural to suppose that the two spin-aligned field-source preons move circularly in orbits where the orbital phase of one preon differs by 180 degrees from the second preon.  This implies that the relativistic velocities of the two spin-aligned preons are equal and opposite, or very close to it, at all times.  This will tend to set up a cancellation of the net inverse-square Coulomb-like magnetic field due to the two spin-aligned preons. An alternative explanation for how oppositely-spinning particles might interact coherently and in accordance with Eq. (\ref{modfact_retarded}) is therefore needed.

The preon model may perhaps more plausibly provide a mechanism of relative kinetic energy modulated magnetic interactions of unequal mass particles. This is a second restriction so far on the present model, that is obviously inconsistent with observation.  The wave-mechanical approach to quantum mechanics was initially applied to the system of the hydrogen atom, where the constituents consisting of an electron and proton have significantly different masses.  If the interactions are between a fundamental charged preon of spin one-half and its antiparticle, then the assumption of equal-mass interacting particles and thus equal rest-frame zitter frequencies is not violated.  In the model, the quarks that compose the proton are each composed of three spin-half preons, and where the charged preon is of only one type (along with its anti-particle).  This makes plausible that the interaction as described here might also occur between an electron and a proton.

Returning to the problem of the interaction of oppositely-spinning electrons each modeled as a single point charge in relativistic circulatory motion, it seems worth considering that a spinning top observed in a mirror appears to have an opposite direction of spin.  Thus, the magnetic field caused by a spinning electron can interact coherently with an oppositely-spinning particle if it is first reflected by a perfectly conducting plane surface.  Obviously no such reflecting surface exists within an atom or at an atomic scale, but the property of a mirror image reversing the apparent direction of a rotation is independent of the details of the mirror geometry, and applies as well to the case of a spherical mirror.  If the test zitter particle is envisioned as being at the center of a large spherical perfectly reflecting surface, then the image of the source zitter particle would appear to an observer at the test particle location to be rotating oppositely to the actual particle.  To a significant extent, the distant absorber of the absorber theory of Wheeler and Feynman \cite{WheelerFeynman1945} behaves like a perfectly reflecting smooth surface, and because in absorber theory a radiating particle is influenced in simultaneity with its acceleration causing the acceleration, because a distant accelerated absorbing charge re-radiates toward the original source along the time-advanced path.   Although the center of focus of the radiation returned from the absorber is at the radiating charge, there is no reason that a charge nearby the radiation source will not be influenced by it as well.  However if the source is undergoing a circulatory motion, the direction of the circulation will appear to be opposite the actual source in the mirror image.  Thus an interaction between oppositely-oriented spinning zitter particles in accordance Eq. (\ref{modfact_retarded}) seems plausible if Wheeler-Feyman absorber theory (WFAT) is correct. However, it must be noted that the absorber response as found in WFAT only loosely resembles a spherical mirrow. The quantitative similarity is developed below as needed to incorporate Osiak relativity into WFAT. It can also be noted that WFAT and the present model share a common postulate that time-advanced radiation exists and has similar influence on charged matter as does time-retarded radiation.

\subsection{Motion Induced Modulation of Time-Symmetric Interaction}
\label{ModulationSymmetric}

It is also shown in \cite{LushY16} how a magnetic force component with a de Broglie wavelength modulation period can be obtained for low-velocity radial relative motion between the zitter particles, if a time-symmetric electromagnetic interaction is assumed, where the total electromagnetic field is a combination of advanced as well as retarded fields. A combination of advanced and retarded electromagnetic forces was considered by Schild \cite{Schild1963} in the interaction of two classical point-like charged particles and by Schild and Schlosser \cite{SchildSchlosser1965} for the case of particles with intrinsic spin and magnetic moment. However, these efforts did not attribute the spin to relativistic circulatory point-charge motion as does the present model.   

The modulating factor on the radial inverse-square law magnetic force for the test zitter particle moving in the advanced field of the source zitter particle, equivalent to Eq. (\ref{retarded_modfact_tret}) for the retarded interaction, is
\begin{eqnarray}
 \mbox{\boldmath$\beta$}_t \cdot   \mbox{\boldmath$\beta$}_{s,{\textnormal{{\scriptsize adv}}}}(t) \equiv {\cal M}_{{\textnormal{{\scriptsize adv}}}}(t) \nonumber \\ = \cos\left[\gamma \omega_z( t - \mbox{\boldmath$\upsilon$} \cdot \mbox{\boldmath$r$}/c^2) - \omega_z t_{\textnormal{{\scriptsize adv}}} - \phi_0\right], 
\end{eqnarray}
where \(t_{\textnormal{{\scriptsize adv}}} =  t + r_{\textnormal{{\scriptsize adv}}}/c \) is the advanced time, with \(r_{\textnormal{{\scriptsize adv}}}/c\) the  displacement from the test zitter particle charge at the present time \(t\) to the field-source zitter particle center of motion at the future time when the magnitude of the four-displacement is null. With the source zitter particle assumed stationary, \(r_{\textnormal{{\scriptsize adv}}}(t) = r_{\textnormal{{\scriptsize ret}}}(t) = r(t) \equiv r\), and so  \begin{equation}
{\cal M}_{{\textnormal{{\scriptsize adv}}}}(t) = \cos ( \omega_z \left[ (\gamma - 1) t - \gamma  \mbox{\boldmath$\upsilon$} \cdot \mbox{\boldmath$r$}/c^2 -  r/c\right]  - \phi_0).
\label{modfact_advanced} 
\end{equation}

The modulating factor on the radial inverse-square  magnetic force for the test zitter particle moving in the sum of the retarded and advanced fields is then (using the identity \(\cos\theta + \cos\phi = 2\cos((\theta + \phi)/2)\cos((\theta - \phi)/2) \)),
\begin{eqnarray}
 \mbox{\boldmath$\beta$}_t \cdot \mbox{\boldmath$\beta$}_{s,{\textnormal{{\scriptsize ret}}}} + \mbox{\boldmath$\beta$}_t \cdot   \mbox{\boldmath$\beta$}_{s,{\textnormal{{\scriptsize adv}}}} = {\cal M}_{{\textnormal{{\scriptsize ret}}}}(t)  +  {\cal M}_{{\textnormal{{\scriptsize adv}}}}(t) \nonumber \\ \equiv {\cal M}_{{\textnormal{{\scriptsize sym}}}}(t) = {\cal M}_{{\textnormal{{\scriptsize sym}}}}(\mbox{\boldmath$r$},t) \nonumber \\ = 2 \cos \left(\omega_z \left[ (\gamma - 1) t - \gamma  \mbox{\boldmath$\upsilon$} \cdot \mbox{\boldmath$r$}/c^2\right]\right) \nonumber \\ \times \cos \left(\omega_z r/c + \phi_0\right). 
\label{SSTimeSymmetricFactor}
\end{eqnarray}

It is thus possible in the time-symmetric case to factor the modulation of the radial magnetic force into one sinusoidal factor with frequency depending on the test particle velocity and independent of the interparticle separation, and another with frequency explicitly dependent on the separation. Such factorization is not possible for the retarded-only interaction, as illustrated by Eq. (\ref{modfact_retarded}).

In the case of nonradial motion of the test zitter particle, Eq. (\ref{SSTimeSymmetricFactor}) reproduces the modulation behavior of the retarded-only interaction, where the modulation spatial period is the de Broglie wavelength for nonrelativistic velocity, but approaches half the de Broglie wavelength in the luminal limit.

\section{Similarity of the Time-Symmetric Force Modulation to the Schr\"odinger Wavefunction}
\label{Implications}

To investigate further what relationship the inverse-square radial magnetic force and its modulation might have to quantum behavior, its possible similarity  will be considered in this section to the Schr\"odinger wavefunction.  Schr\"odinger developed a nonrelativistic  wave equation for a scalar field based on de Broglie's matter wave.  Schr\"odinger recognized that the de Broglie wave  group velocity equality with particle velocity could be used to link the matter wave spatial derivative to the particle momentum. Along with a time variation relatable to the particle energy, this enabled construction of a Hamiltonian using partial derivatives of a ``wavefunction'' that is a complex extension of a nonrelativistic approximation of the de Broglie wave. 

In the interacting zitter particle model, the Coulomb-like radial magnetic force modulation can be factored such that one factor satisfies a partial differential equation that is similar to the free-particle Schr\"odinger equation (SE), where the occurrences of \(\hbar\) in the SE are replaced by the quantity \(s\), the intrinsic angular momentum of the zitter particle. Thus, for the electron, occurrences of \(\hbar\) in the Schr\"odinger equation are replaced by \(s=\hbar/2\).

\subsection{Similarity of the Time-Symmetric Force Modulation to the de Broglie Matter Wave}
\label{Sim2MW}

The modulating factor due to the time-symmetric field was found in Eq. (\ref{SSTimeSymmetricFactor})  as \({\cal M}_{{\textnormal{{\scriptsize sym}}}} = {\cal M}_{{\textnormal{{\scriptsize ret}}}} + {\cal M}_{{\textnormal{{\scriptsize adv}}}}\).  Dropping the leading factor of \(2\) as irrelevant here let
\begin{eqnarray}
{\cal M}_{{\textnormal{{\scriptsize sym}}}}(x,t)  = \nonumber \\ 
\cos \left(\omega_z \left[ (\gamma - 1) t - \gamma  \mbox{\boldmath$\upsilon$} \cdot \mbox{\boldmath$r$}/c^2\right]\right) \cos \left(\omega_z r/c + \phi_0\right) \nonumber \\ \equiv {\cal M}_1 {\cal M}_2. 
\label{SSTimeSymmetricFactor3}
\end{eqnarray}

The space and time varying phase of the first cosine factor \({\cal M}_1\) of Eq. (\ref{SSTimeSymmetricFactor3}) can be written as
\begin{equation}
\omega_z \left[ (\gamma - 1) t - \gamma  \upsilon \hat{\mbox{\boldmath$\upsilon$}} \cdot \mbox{\boldmath$r$}/c^2\right] = \omega t - kd,
\end{equation}
where \(d = \hat{\mbox{\boldmath$\upsilon$}} \cdot \mbox{\boldmath$r$}\),  wavenumber \(k = \gamma \omega_z \upsilon/c^2 = \gamma \omega_z \beta/c\) and frequency \(\omega = \omega_z (\gamma -1)\). The phase velocity is 
\begin{equation}
\upsilon_p = \frac{\omega}{k} = \frac{\omega_z (\gamma -1) c}{\gamma \omega_z \beta} = \frac{c(\gamma -1)}{\gamma\beta}.
\label{PhaseVel}
\end{equation}

For \(\beta\) near unity, the phase velocity is thus superluminal and approaches that of the de Broglie matter wave. For small \(\beta\), \(\gamma \approx 1 + {\beta}^2/2\) and (unlike the de Broglie matter wave),  
\begin{equation}
\upsilon_p \approx \frac{\omega_z (\beta^2/2) c}{ \omega_z \beta} = \frac{c\beta}{2} = \frac{\upsilon}{2}.
\end{equation}

For the spin-half zitter particle with \(\omega_z = 2 \omega_B = 4 \pi c / \lambda_0\), the wavelength \(\lambda = 2 \pi/ k\) is 
\begin{equation}
\lambda = \frac{2 \pi c}{ \gamma\beta\omega_z} = \frac{\lambda_0}{2\gamma\beta}= \frac{h}{2\gamma m_0 c \beta} = \frac{h}{2\gamma m_0 \upsilon}= \frac{h}{2 p},
\end{equation}
with \(p = \gamma m_0 \upsilon\) the test zitter particle momentum.  The  wavelength of the first modulating factor of Eq. (\ref{SSTimeSymmetricFactor3}) for the spin-half zitter particle model of the Dirac electron is thus half the de Broglie wavelength. The wavelength  for the spin-one zitter particle is equal to the de Broglie wavelength.  

The group velocity of a wave of frequency \(\omega\) and wavenumber \(k\) is generally \(\upsilon_g = d\omega/dk \).
Following de Broglie, and with \(\omega = \omega_z(\gamma - 1)\) and \(k = \gamma \omega_z \beta/c\), the group velocity of the first cosine factor of Eq. (\ref{SSTimeSymmetricFactor3}) is thus calculated as
\begin{eqnarray}
\upsilon_g = \frac{d\omega}{dk} = \frac{d\omega}{d\beta}\frac{d\beta}{dk} = \frac{d\omega}{d\beta} \left[\frac{dk}{d\beta}\right]^{-1}\nonumber \\ = \frac{\omega_z c}{\omega_z} \frac{d(\gamma -1)}{d\beta} \left[\frac{d(\gamma \beta)}{d\beta}\right]^{-1}
 = c \frac{d\gamma}{d\beta} \left[ \gamma + \beta\frac{d\gamma}{d\beta}\right]^{-1}.
\label{groupvel}
\end{eqnarray}
Substituting for \(d\gamma/d\beta = \gamma^3 \beta \), and with\( \gamma^2 \beta^2 = \gamma^2 - 1\),
\begin{equation}
\upsilon_g =  \gamma^3 \beta c  \left[ \gamma + \gamma^3 \beta^2 \right]^{-1} = \frac{c\gamma^2 \beta}{1 + \gamma^2 \beta^2} =  \upsilon. 
\end{equation}
The group velocity of the first cosine factor of Eq. (\ref{SSTimeSymmetricFactor3}) is therefore, like the group velocity of the de Broglie matter wave, equal to the velocity of the particle containing the oscillating dipole. 

The spatial period is the distance the test zitter particle travels during one cycle of modulation. The spatial period \(D_2\) of the second factor, \({\cal M}_2\), on the right hand side of Eq. (\ref{SSTimeSymmetricFactor3}), that depends on the interparticle separation, may be found as
\begin{equation}
\omega_z \left[\frac{D_2}{c}\right] = 2 \pi,
\end{equation}
so
\begin{equation}
D_2 = \frac{2 \pi c}{\omega_z} = \frac{2 \pi \hbar c}{2  m_e c^2 }  = \frac{h}{2 m_e c } = \frac{\lambda_0}{2}.
\end{equation}
The second factor on the right hand side of Eq. (\ref{SSTimeSymmetricFactor3}) thus oscillates with a spatial period that is half the Compton wavelength.

\subsection{Partial Differential Equation of Constraint for Magnetically-Interacting Dirac Particles}

In the following, for simplicity, the case of one spatial dimension is considered.  All of the results generalize straightforwardly to three dimensions.  From Eq. (\ref{SSTimeSymmetricFactor3}), and restricting to one spatial dimension, let 
\begin{eqnarray}
{\cal M}_1(x,t) \equiv \cos\left( \omega_z \left[(\gamma - 1) t - \frac{\gamma\beta x}{c}\right]\right) \nonumber \\ = \cos\left(\omega_z \left[\frac{\gamma\beta x}{c} - (\gamma - 1) t\right]\right),
\label{MSym_avg} 
\end{eqnarray}
where \(t\) here is again the time coordinate in the field-source zitter particle rest frame, and \(x\) is the separation between the test and source zitter particles in the source zitter particle rest frame.

With \(\omega_z = m_0 c^2 / s = E_0 / s\) with \(s\) the intrinsic spin, and \(K \equiv E - E_0 = (\gamma - 1)E_0\) the kinetic energy, Eq. (\ref{MSym_avg}) becomes
\begin{eqnarray}
{\cal M}_1(x,t) = \cos\left(\frac{E_0}{s} \left[\frac{\gamma\beta x}{c} - (\gamma - 1) t\right]\right)\nonumber \\ = \cos\left(\frac{1}{s} \left[\gamma m_0 \upsilon x - K t \right]\right).
\label{MSym_avg_gen_s} 
\end{eqnarray}

For the spin-half zitter particle, \(s = \hbar / 2\), and with \(p = \gamma m_0 \upsilon\) the relativistic momentum,
\begin{equation}
{\cal M}_1(x,t) = \cos\left(\frac{2}{\hbar} \left[p x - K t \right]\right).
\label{MSym_avg_spin_half} 
\end{equation}

Noting that \(e^{i\theta} = \cos \theta + i \sin \theta\), where \(i\) is the imaginary unit, the modulating factor given by Eq. (\ref{MSym_avg_spin_half}) can be justifiably redefined  as 
\begin{equation}
{\cal M}(x,t) = \exp \left(i\frac{2}{\hbar} \left[p x - K t \right]\right),
\label{MSym_avg_complex} 
\end{equation}
with the understanding that only the real part of expressions involving it need have physical significance.  Then,  
\begin{equation}
\frac{\partial {\cal M}(x,t)}{\partial t} = -i\frac{2 K}{\hbar} {\cal M}(x,t),
\end{equation}
and
\begin{equation}
\frac{\partial^2 {\cal M}(x,t)}{\partial x^2}  = -\frac{4p^2}{\hbar^2} {\cal M}(x,t).
\end{equation}

If the time-dependent Schr\"odinger wavefunction \(\Psi(x,t)\) is replaced by \( {\cal M}(x,t)\) in the one-dimensional free-particle Schr\"odinger equation as 
\begin{equation}
\frac{-\hbar^2}{2m}\frac{\partial^2}{\partial x^2} {\cal M}(x,t) = i \hbar \frac{\partial}{\partial t} {\cal M}(x,t),
\end{equation}
then the result is
\begin{equation}
4\frac{p^2}{2m} = 2K,
\end{equation}
in contradiction with the kinetic energy relation to the momentum as \(K = p^2/2m\) in the non-relativistic limit. However, \({\cal M}(x,t)\) satisfies
\begin{equation}
 \frac{-s^2}{2m_0}\frac{\partial^2}{\partial x^2} {\cal M}(x,t) = i s \frac{\partial}{\partial t} {\cal M}(x,t),
\label{hbar2s}
\end{equation}
where \(s\) is the spin of the zitter particle, with \(s=\hbar/2\) for the Dirac electron.  This equation is formally identical to the free-particle time-dependent Schr\"odinger equation (SE) in one dimension if \(\hbar\) is replaced by the electron spin \(s\). For the electron Eq. (\ref{hbar2s}) thus becomes
\begin{equation}
\frac{-\hbar^2}{2 m_0}\frac{\partial^2}{\partial x^2} {\cal M}(x,t) = i 2\hbar \frac{\partial}{\partial t} {\cal M}(x,t).
\label{TDSlikeEq}
\end{equation}
This result is formally identical to the time-dependent free-particle SE apart from the factor of \(2\) on the right-hand side.  

A time-dependent equation involving a potential function, similarly to the time-dependent SE, does not exist within the present model. This is a consequence of the fact that in the case of magnetically-interacting zitter particles, the time frequency of the force modulation is proportional to the kinetic energy \(K\) rather than the total energy \(E = K + V\). 

It is of interest to separate variables so as to obtain a time-independent equation. Letting 
\begin{eqnarray}
{\cal M}(x,t) = \exp\left(i2 \left[p x - Kt \right]/\hbar\right) = M(x)f(t),
\label{MSym_separated} 
\end{eqnarray}
where \(M(x) = \exp\left(i2 p x /\hbar\right)\), and \(f(t) = \exp\left(-i2 Kt /\hbar\right)\), then
\begin{equation}
\frac{\partial^2 M(x)}{\partial x^2} = -\frac{4 p^2}{\hbar^2} M(x),
\end{equation}
\begin{equation}
\frac{\partial f(t)}{\partial t} = -i\frac{2 K}{\hbar} f(t).
\end{equation}
Also,
\begin{equation}
\frac{\partial M(x)}{\partial t} = \frac{\partial f(t)}{\partial x} = 0.
\end{equation}
Eq. (\ref{TDSlikeEq}) then becomes
\begin{equation}
-f(t)\frac{\hbar^2}{2 m_0}\frac{\partial^2}{\partial x^2} M(x) = i 2\hbar M(x) \frac{\partial}{\partial t}f(t).
\label{TDSlikeEqSep1}
\end{equation}
Evaluating the derivatives yields
\begin{equation}
f(t)\frac{\hbar^2}{2 m_0}\frac{4 p^2}{\hbar^2} M(x) = -i 2\hbar M(x) i\frac{2 K}{\hbar} f(t),
\label{TDSlikeEqSep2}
\end{equation}
which reduces to \(p^2/2 m_0 = K\), the low-velocity approximation for the kinetic energy. Returning to Eq. (\ref{TDSlikeEqSep1}), evaluating the derivative on the right-hand side only, and dividing out \(f(t)\)) on both sides obtains
\begin{equation}
-\frac{\hbar^2}{2 m_0}\frac{\partial^2}{\partial x^2} M(x) = 4 K M(x).
\label{TDSlikeEqSep3}
\end{equation}

With \(K = E - V\), where \(V = V(x)\) is a potential function of position, 
\begin{equation}
-\frac{\hbar^2}{2 m_0}\frac{\partial^2}{\partial x^2} M(x) + 4 V M(x) = 4 E M(x).
\label{TDSlikeEqSep4}
\end{equation}
This result is formally similar to the time-independent Schr\"odinger equation in one dimension, but the two factors of \(4\) are inconsistent with observation.

\section{Time-Symmetric Magnetic Force Modulation Assuming Osiak Relativity}
\label{TSymFMOsiak}

Based on the relativistic law of motion attributed to Minkowski as \(F = m_0 A = m_0 dU/d\tau =  m_0 \gamma dU/dt\), where \(F\) is the four-force, \(m_0\) is invariant mass, \(A\) is the four-acceleration, \(U\) is four-velocity, and \(\tau\) is the proper time, Osiak \cite{OsiakY19} derives alternative formulas for the relativistic rest and total energies as \(E_0 = m_0 c^2/2\) and \(E = m_0 \gamma^2 c^2/2\), and relativistic kinetic energy \( K = E - E_0 = m_0 \gamma^2 \upsilon^2/2 \). It's worth noting here that in the Osiak formulation, the relativistic kinetic energy is exactly \(p^2/2m_0\), where \(p\) is the relativistic momentum. This leads to the time component \(p_0 = \gamma m c\) of the four-momentum equatable to \(\sqrt{2 m_0 E}\) in Osiak relativity, as opposed to \(p_0 = E/c\) in Einsteinian relativity. \footnotemark[1]

\footnotetext[1]{It is of relevance to the Dirac relativistic electron theory that the time component of the four-momentum, \(p_0 = \gamma m c = E/c\) in standard relativity becomes \(p_0 = \gamma m c = \sqrt{2mE}\) in Osiak relativity. This allows for negative values of the time component of four-momentum without introducing the concept of negative energy.}

\subsection{De Broglie Wave Properties Assuming Osiak Relativity}
\label{DBWOsiak}

Before considering how the time-symmetric magnetic force modulation and its similarity to the Schr\"odinger wavefuntion are affected by an assumption that Osiak's revision of special relativity is correct, it is of interest to consider initially the de Broglie matter wave.  The de Broglie frequency \(\omega_B  = E_0/\hbar = m_0 c^2/\hbar\) as given previously by Eq. (\ref{debroglieclock}) will be halved by modifying the rest energy of a particle of invariant mass \(m_0\) to become  \(\omega = E'_0/\hbar = m_0 c^2/2\hbar = \omega_B / 2\), with \(E'_0 \equiv m_0 c^2 /2\) the rest energy assuming Osiak relativity is correct. However, if it is also considered that de Broglie based his wave frequency on the intrinsic angular momentum \(\hbar\) of the photon, without knowledge in 1923 that the electron intrinsic angular momentum \(\hbar/2 \) is half that of a photon, it is natural to rewrite the de Broglie frequency based on Osiak relativity and taking into account the electron spin as  

\begin{equation}
\omega'_B  \equiv \frac{E'_0}{s} = \frac{m_0 c^2}{2}\frac{2}{\hbar} = \frac{m_0 c^2}{\hbar} = \omega_B. 
\label{deBroglieClockOsiak}
\end{equation}  
That is, the net result of halving both the rest energy, per Osiak relativity, and the intrinsic angular momentum of the particle, as needed to describe the electron rather than the photon, is a wave frequency equal to that hypothesized by de Broglie.  

With the de Broglie frequency also derivable based on the electron intrinsic angular momentum combined with the Minkowski-Osiak mass-energy equivalency, the remainder of de Broglie's argument concerning phase harmony between of moving particle with an internal oscillation and a wave of superluminal phase velocity \(\upsilon_p = c/\beta\) is unaffected by the alternative rest energy. The resulting wavelength remains \(\lambda = h / p\), with \(p=\gamma m \upsilon\) the relativistic momentum. The de Broglie matter wave can thus coexist with Osiak's version of special relativity, provided that the photon intrinsic angular momentum \(\hbar\) is  replaced by the electron spin \(\hbar/2\). This seems a reasonable circumstance for a matter wave associated with the electron rather than the photon.

\subsection{Time-Symmetric Force Modulation for Osiak Relativity}
\label{Sim2MW}

Assuming rest energy \(E_0 = m_0 c^2/2\), the first factor of Eq. (\ref{SSTimeSymmetricFactor3}) for the time-symmetric Coulomb-like magnetic force modulation, after restricting to one spatial dimension, becomes
\begin{equation}
{\cal M}_1(x,t) = \cos\left(\omega_z \left[\frac{\gamma\beta x}{c} - (\gamma - 1) t\right]\right),
\label{MSym1_osiak} 
\end{equation}
where \(\omega_z = E_0 / s =  m_0 c^2 / (2s) =  m_0 c^2 / \hbar\), here,  with \(s\) the spin.

The space and time varying phase of Eq. (\ref{MSym1_osiak}) can be written as
\begin{equation}
\omega_z \left[ (\gamma - 1) t - \gamma  \upsilon \hat{\mbox{\boldmath$\upsilon$}} \cdot \mbox{\boldmath$r$}/c^2\right] = \omega t - kd,
\end{equation}
where \(d = x\), wavenumber \(k= \gamma \omega_z \beta/c\) and frequency \(\omega = \omega_z (\gamma -1)\). The phase velocity is 
\begin{equation}
\upsilon_p = \frac{\omega}{k} = \frac{\omega_z (\gamma -1) c}{\gamma \omega_z \beta} = \frac{c(\gamma -1)}{\gamma\beta}. 
\end{equation}
This differs from the de Broglie wave phase velocity \(\upsilon_B = c/\beta = c^2/\upsilon \), but is identical to the expression found previously (i.e., Eq. (\ref{PhaseVel})) based on the Einstein mass-energy equivalency \(E=mc^2\).

For the spin-half zitter particle with (in the Minkowski-Osiak case) \(\omega_z = \omega_B = m_0 c^2  / \hbar\), and wavenumber \(k= \gamma \omega_z \beta/c\), the wavelength \(\lambda = 2 \pi/ k\) is 
\begin{equation}
\lambda = \frac{2 \pi c}{ \gamma\beta\omega_z} = \frac{2 \pi c \hbar}{ \gamma\beta m_0 c^2} = \frac{h}{\gamma m_0 c \beta} = \frac{h}{\gamma m_0 \upsilon} = \frac{h}{p},
\end{equation}
with \(p = \gamma m_0 \upsilon\) the test zitter particle momentum.  The  wavelength of the first modulating factor of Eq. (\ref{SSTimeSymmetricFactor3}) for the spin-half zitter particle model of the Dirac electron is thus the de Broglie wavelength.

The group velocity of the first cosine factor of Eq. (\ref{SSTimeSymmetricFactor3}) calculated using Einstein-Osiak relativity arrives again at the result \(\upsilon_g = \upsilon\). This can be seen by inspection of Eq. (\ref{groupvel}) determining the group velocity of the position and time dependent modulation factor on the Coulomb-like time-symmetric magnetic force assuming the Einstein mass-energy equivalency.  The appearance of \(\omega_z\) in both the numerator and denominator of the expression for the group velocity shows that halving \(\omega_z\) has no effect on the result.

The spatial period \(D_2\) of the second factor on the right hand side of Eq. (\ref{SSTimeSymmetricFactor3}), that depends on the interparticle separation, is found as
\begin{equation}
D_2 = \frac{2 \pi c}{\omega_z} = \frac{2 \pi \hbar c}{ m_e c^2 }  = \frac{h}{ m_e c } = \lambda_0.
\end{equation}
Using Einstein-Osiak relativity, the second factor on the right hand side of Eq. (\ref{SSTimeSymmetricFactor3}) thus oscillates with a spatial period that is equal the Compton wavelength.

\subsection{Partial Differential Equation of Constraint using Osiak's Relativistic Energy}

For the spin-half zitter particle with spin \(s = \hbar/2\), and with rest energy \(E_0 = m_0 c^2 / 2\) according to Osiak's special relativity energy calculation, then \(\omega_z = E_0 / s =  m_0 c^2 / \hbar = \omega_B\), where the de Broglie frequency  \(\omega_B\) was defined in Eq. (\ref{debroglieclock}). With \({\cal M}_1(x,t)\) incorporating Osiak relativity as according to Eq. (\ref{MSym1_osiak}),
\begin{eqnarray}
\frac{\partial^2{\cal M}_1(x,t)}{\partial x^2} = -\left(\frac{\omega_B\gamma\beta}{c}\right)^2{\cal M}_1(x,t) \nonumber \\ = -\left(\frac{m_0 c^2 \gamma\beta}{\hbar c}\right)^2{\cal M}_1(x,t) \nonumber \\ = -\left(\frac{p}{\hbar }\right)^2{\cal M}_1(x,t),
\end{eqnarray}
with \(p = \gamma m_0 \upsilon\).

Also,  (with \(\omega_z =  m_0 c^2 / \hbar \equiv \omega_B\) here),  \(\omega_z(\gamma - 1) =  m_0 (\gamma - 1) c^2 / \hbar\). In the low-velocity approximation, and in Osiak relativity, \( m_0 (\gamma - 1) c^2 / \hbar \approx m_0 (\beta^2/2) c^2 / \hbar  = K / \hbar\), and Eq. (\ref{MSym1_osiak}) becomes, for non-relativistic velocity,

\begin{equation}
{\cal M}_1(x,t) = \cos\left(\frac{1}{\hbar} \left[p x - K t \right]\right).
\label{MSym1_osiak_2} 
\end{equation}

Using the complex extension of the  modulating factor given by Eq. (\ref{MSym1_osiak_2}),
\begin{equation}
{\cal M}(x,t) = \exp \left(\frac{i}{\hbar} \left[p x - K t \right]\right),
\label{MSym1_complex_osiak} 
\end{equation}
and \(K = p^2/2m\), then \({\cal M}(x,t)\) satisfies, in the non-relativitic approximation, 
\begin{equation}
\frac{-\hbar^2}{2 m_0}\frac{\partial^2}{\partial x^2} {\cal M}(x,t) = i \hbar \frac{\partial}{\partial t} {\cal M}(x,t),
\label{TDSlikeEq_osiak}
\end{equation}
 which is formally identical to the time-dependent free-particle Schr\"odinger equation.  Also, like the Schr\"odinger equation, it is valid only non-relativistically (here because in Osiak relativity, the kinetic energy is \(E_k = m_0 (\gamma ^2 - 1) c^2/2 = p^2/2m = K\), but \(\gamma\) only appears to the first power in the \((\gamma-1)t\) part of the exponential argument of the right-hand side of (\ref{MSym1_osiak}), as it arises from a Lorentz trasformation).  However, it can be shown \cite{LushY22} that Osiak relativity is fully compatible with Dirac's relativistic quantum theory of the electron.

\subsection{Separation of Variables of Partial Differential Equation of Constraint - Osiak's Relativistic Energy}
\label{SepVarOsiak}

The modulation factor \({\cal M}(x,t)\) of Eq. (\ref{MSym1_complex_osiak}) may be factored as
\begin{eqnarray}
{\cal M}(x,t) = f(t)M(x) \nonumber \\ = \exp\left[-i2\left(K t \right)/\hbar\right] \exp \left[i\left(p x \right)/\hbar\right].
\end{eqnarray}
Eq. (\ref{TDSlikeEq_osiak}) then becomes
\begin{equation}
f(t)\frac{-\hbar^2}{2 m_0}\frac{\partial^2}{\partial x^2} M(x) = i M(x) \frac{\hbar}{2} \frac{\partial}{\partial t} f(t).
\end{equation}
Evaluating the derivatives,
\begin{eqnarray}
f(t)\frac{-\hbar^2}{2 m_0}\frac{\partial^2}{\partial x^2} M(x) = i M(x) \frac{\hbar}{2} \left[\frac{-i2 K }{\hbar}\right] f(t) \nonumber \\ = K M(x) f(t).
\end{eqnarray}
Dividing by \(f(t)\) obtains
\begin{equation}
\frac{-\hbar^2}{2 m_0}\frac{\partial^2}{\partial x^2} M(x) = K M(x). 
\end{equation}

With total energy \(E = K(x) + V(x)\), where \(V(x)\) is the potential energy,
\begin{equation}
\frac{-\hbar^2}{2 m_0}\frac{\partial^2}{\partial x^2} M(x) + V(x) M(x) = E M(x). 
\label{TIndSlikeEq_osiak}
\end{equation}

Osiak's interpretation of energy in special relativity thus leads to an equation of motion of the modulation factor on the Coulomb-like magnetic force that is formally identical to the time-independent Schr\"odinger equation.

\subsection{Imaginary Character of Advanced Electromagnetic Field in Osiak Relativity}

Osiak's proposed form of special relativity implies that the electromagnetic field is a complex quantity, due to the allowance of a negative temporal component of four-momentum.  From Jackson Eq. (12.114),

\begin{equation}
\Theta^{00} =\frac{1}{8\pi}\left( E^2 + B^2   \right)
\end{equation}

and

\begin{equation}
\Theta^{0i} =\frac{1}{4\pi}\left( \mbox{\boldmath$E$}  \times \mbox{\boldmath$B$} \right)_i
\label{PoyntingVector}
\end{equation}

are components of the covariant symmetric electromagnetic stress tensor that correspond to components of  the four-momentum density of the electromagnetic field.   Allowing the temporal component of electromagnetic four-momentum to be negative in the case of time-advanced electromagnetic fields leads to the expectation that advanced electric and magnetic field are pure imaginary quanities, that is, 

\begin{equation}
\Theta^{00}_{{\textnormal{{\scriptsize adv}}}} =\frac{1}{8\pi}\left( E_{{\textnormal{{\scriptsize adv}}}}^2 + B_{{\textnormal{{\scriptsize adv}}}}^2   \right) \le 0
\end{equation}

Since either the electric or magnetic field may vanish separately, the advanced fields must be pure imaginary.  The momentum transported by the advanced field according to (\ref{PoyntingVector}) is thus real.  Also, the direction of energy transport by the electromagnetic field is reversed for time-advanced radiation. 

Creation of an imaginary field requires either an imaginary charge, or charge motion in an imaginary space.  Here it will be supposed that charge associated with particles of negative Lorentz factor is imaginary, and that all positive charge is imaginary, while negative charge is real, for forward-time observers.  

\subsection{Complex Form of Modulation Factor Naively Expected Due to Osiak Relativity}

The modulating factor on the radial inverse-square law magnetic force for the test zitter particle moving in the advanced field of the source zitter particle, equivalent to Eq. (\ref{retarded_modfact_tret}) for the retarded interaction, aussuming advanced fields are imaginary as predicted by Osiak relativity, is
\begin{eqnarray}
{\cal M}_{{\textnormal{{\scriptsize adv}}}}(t) \equiv i\mbox{\boldmath$\beta$}_t \cdot   \mbox{\boldmath$\beta$}_{s,{\textnormal{{\scriptsize adv}}}}(t) \nonumber \\ = i\cos\left[\gamma \omega_z( t - \mbox{\boldmath$\upsilon$} \cdot \mbox{\boldmath$r$}/c^2) - \omega_z t_{\textnormal{{\scriptsize adv}}} - \phi_0\right], 
\end{eqnarray}
where \(t_{\textnormal{{\scriptsize adv}}} =  t + r_{\textnormal{{\scriptsize adv}}}/c \) is the advanced time, with \(r_{\textnormal{{\scriptsize adv}}}/c\) the  displacement from the test zitter particle charge at the present time \(t\) to the field-source zitter particle center of motion at the future time when the magnitude of the four-displacement is null. With the source zitter particle assumed stationary, \(r_{\textnormal{{\scriptsize adv}}}(t) = r_{\textnormal{{\scriptsize ret}}}(t) = r(t) \equiv r\), and so  
\begin{equation}
{\cal M}_{{\textnormal{{\scriptsize adv}}}}(t) = i\cos ( \omega_z \left[ (\gamma - 1) t - \gamma  \mbox{\boldmath$\upsilon$} \cdot \mbox{\boldmath$r$}/c^2 -  r/c\right]  - \phi_0).
\label{modfact_advanced_Osiak} 
\end{equation}

The modulating factor on the radial inverse-square  magnetic force for the test zitter particle moving in the sum of the retarded and advanced fields is then 

\begin{eqnarray}
{\cal M}_{{\textnormal{{\scriptsize sym}}}}(t) \equiv \mbox{\boldmath$\beta$}_t \cdot \mbox{\boldmath$\beta$}_{s,{\textnormal{{\scriptsize ret}}}} + i\mbox{\boldmath$\beta$}_t \cdot   \mbox{\boldmath$\beta$}_{s,{\textnormal{{\scriptsize adv}}}}  \nonumber \\ = {\cal M}_{{\textnormal{{\scriptsize ret}}}}(t)  +  {\cal M}_{{\textnormal{{\scriptsize adv}}}}(t)  = {\cal M}_{{\textnormal{{\scriptsize sym}}}}(\mbox{\boldmath$r$},t) \nonumber \\ = \cos ( \omega_z \left[ (\gamma - 1) t - \gamma  \mbox{\boldmath$\upsilon$} \cdot \mbox{\boldmath$r$}/c^2 +  r/c\right]  - \phi_0)\nonumber \\ +  i\cos ( \omega_z \left[ (\gamma - 1) t - \gamma  \mbox{\boldmath$\upsilon$} \cdot \mbox{\boldmath$r$}/c^2 -  r/c\right]  - \phi_0) 
\label{naive_mod_fact}
\end{eqnarray}

Such a modulation is not obviously useful, as it does not display the similarity to the Schroedinger wavefunction found under the assumption of a real advanced field. It cannot be written as a product of two factors, where one of the factors solves the Schroedinger equation (and the other separates out).  It will be seen that Eq. (\ref{naive_mod_fact}) is incomplete due to lack of accounting for absorption and re-radiation as described in Wheeler-Feynman Absorber Theory (WFAT), however.  To arrive at the complete expression, it is necessary to modify and extend WFAT on account of the complexity of the EM field in Osiak relativity.

\section{Necessary Extension of Wheeler-Feynman Absorber Theory Due to Osiak Relativity}

As noted above, the present model obtains a modulation factor resembling the Schroedinger wavefunction only for the case of aligned spins. Also, it was suggested above that a Schroedinger wavefuction-like modulation factor can be obtained for opposing spins using WFAT, because advanced radiation coherently converging on the source particle will also converge coherently on a nearby test particle, but may be inverted or in-phase depending on whether the absorber advanced radiation is convergent on, or diverging from, the source particle.

Osiak relativity applied to WFAT implies that WFAT can work in a time-symmetric fashion it doesn't exhibit when based on Einstein relativity.  In the original version, WFAT predicts radiative damping due to absorption.  Because advanced radiation from the absorber cancels that from the source, there can be no absorbtion of the advanced field.

Osiak relativity leaves intact the retarded and advanced electromagnetic wave equation solutions for the real EM field, while allowing a second set of solutions for the imaginary EM field.  For the real case, the retarded field is divergent while the advanced field converges on the source.  The imaginary field is time-reversed, so what appears as convergent to a time-forward observer is divergent to the time-reversed observer.  The time-reversed observer sees WFAT working exactly as it does for the time-forward observer, because the imaginary field cannot interfere with the real field.  Thus, the divergent (to the time-reversed observer) time-reversed field causes as apparently-spontaneous emission (i.e., time-reversed absorption) in the distant past,  causing a real \(\pi/2\) phase-advanced field converging on the source according to the time-forward observer.

\subsection{Real field phase-identical to the imaginary advanced field simultaneous with the retarded field}

Osiak relativity implies that advanced fields that transport energy from the present to the past, if they exist, are imaginary.  It has been found above without consideration of Osiak relativity, that assuming a real advanced field can lead to a modulation that is similar to the Schroedinger wavefunction.  That result will be negated if the advanced field is generally imaginary.  However, assuming a version of WFAT is also true, it is possible that a real field phase-identical to the advanced field can act on the test particle, as follows.

According to a time-reversed observer, the advanced field is a real retarded field, and as such may be completely absorbed by a distant absorber.  This creates radiative fields that are both retarded and advanced.  What is advanced to the time-reversed observer is retarded and thus real to the forward-time observer.  The real retarded field from the absorber eventually passes through the source and becomes part of the total retarded field from the source. Prior to converging on the source, it encounters the test particle, and according to WFAT it is phase-identical to the source advanced field. 

The absorber retarded field from the distant past converging on the source in the present, that is real in the forward-time frame, is also needed to account for the radiative damping force described by WFAT, if Osiak relativity is correct and requires advanced fields to be imaginary. 

\subsection{Extended WFAT and the non-observability of advanced fields}

WFAT if correct explains the origin of radiative damping, by hypothesizing reality of the action of the advanced solutions to the electromagnetic wave equation.  At the same time, everyday experience argues against such reality, in that, for example, transmitted radio signals are always received after they are transmitted, never prior to transmission.  Therefore it seems an essential feature of WFAT that modifications as needed if Osiak relativity is correct preserve this feature of the original.  However, it is clear that the precise mechanism in the original WFAT cannot be preserved in the Osiak relativity version and if it is to support the present model of quantum behavior, as the imaginary character of the future absorber advanced field precludes its reinforcement of the half-retarded source field. Also in the original WFAT, the advanced field from the absorber cancels the half-advanced source field, so explaining the lack of observation of time-advanced action.  

In the present model it will be hypothesized that advanced fields are not observed, at least to date, in that their imaginary character restricts their obvious effects on charged matter to positive charges only.  This hypothesis is motivated by two considerations.  First, Osiak relativity, already requiring that advanced fields that carry energy toward the past are necessarily imaginary, seems to require an imaginary character associated with charges in order for such advanced fields to have a real effect on proper acceleration. Second, the present model, and Osiak relativity separately, seem to be compatible with, and may possibly require, that some version of the Shupe-Harari preon model of elementary particles is true.  

In the Shupe-Harari model, there are only two fundamental particles, and the only charged particle is the ``minus'' particle, with one-third the fundamental charge of the electron.  The other particle, the ``zero,'' is neutral.  Each particle has an antiparticle, and it is uniquely the anti-minus preon that is associated with positive charge.  The direct relevance of the preon model to the present model of quantum behavior as described herein is that the effects described herein occur only between charged spin-half particles with equal mass.  Therefore in order for the present model to be, like quantum mechanics, applicable to all charged matter, it seems to require such a restriction to a single type of charged particle. 

A more fundamental reason to hypothesize that positive charge is imaginary, assuming Osiak relativity is correct, is that if position and velocity are real quantities, while advanced electromagnetic field quantities are imaginary, then only if charge is imaginary can acceleration caused by advanced electromagnetic fields be real.  If all positive charge is associated with antimatter, particularly, but even in the restricted case limited to that antimatter has a negative Lorentz factor, it seems necessary that a hypothetical time-reversed observer should infer physical laws very similar if not identical to the forward-time observer.  For example, if as reasonably expected according to Osiak relativity, the positron is seen as an ordinary, negative charge electron to the time-reversed observer, then a cloud-chamber photograph of a pair-creation event shows what is happening in the time-reversed world.  The positron acceleration which is conjecturally here due only to an advanced magnetic field causes position changes that are changes to real coordinates in the forward-time reference frame.  Thus, to describe the motion while assuming the advanced field is imaginary requires that the positve charge be imaginary, or that the Lorentz force law changes to include an imaginary factor when used to describe time-reversed phenomena.

A complete description of physics in Osiak relativity seems also to require an additional kind of space-time coordinate transformation to the Lorentz transformation.  This follows from an expectation that physical description should be independent of the sign of an observer's rest frame, as assigned by a second observer based on the temporal direction of the first observer. That is, there is no Lorentz transfiormation that takes a particle with a positive Lorentz factor to having a negative Lorentz factor.

In original WFAT, the absorber plays a role of retarding the phase of the source retarded wave by \(\pi/2\) radians, when averaged over complete absorption.  The advanced field re-radiated in response by the absorber  is not affected by the absorber, however.  This seems at least potentially to be a temporal bias toward retarded effects compared to advanced ones.  The lack of interaction on the return trip allows the advanced field from the absorber to both reinforce the source retarded field and cancel the source advanced field. Therefore it is an essential feature of the original WFAT.

If instead the absorber advanced field interacts with the absorber time-symmetrically, then it arrives at the source in phase opposition to the source retarded field, while reinforcing the source half-advanced field to full strength. That is, it is phase advanced by \(\pi/2\) radians, on average, exactly undoing the average delay of absorption.  Due to the imaginary nature of advanced fields, the phase opposition relative to the source half-retarded field is inconsequential because the two exist independently.  Reinforcement of the half-retarded field to full strength is then accomplished in the modified extended WFAT by the retarded field from the distant-past absorber, as is described in detail below.  The same half-retarded field from the distant past absorber also gives rise to radiative damping.

It is also important to the present effort that the exact cancelation of the source advanced field that occurs in Einsteinian WFAT is not present in extended modified WFAT that incorporatess Minkowski-Osiak relativity.  As will be seen, the direct advanced source field is an essential contributor to the present model reconstruction of Schroedinger wavefunction, while the advanced absorber field, which inverts due to passing through the source (, according to a time reversed observer), is essential to reconstructing the Pauli model.

Finally, about the question of observability of advanced electromagnetic fields, the present model asserts that they are being observed in every cloud-chamber photograph showing a positron circulating in a magnetic field.  It is simply not possible to influence ordinary matter directly with time advanced fields, as negative electrical charges don't feel their force.  It seems plausible that the time advancement of such advanced action on positrons or other positively charged latter such as positive ions might be measurable.    

\subsection{Time-Reversed Observer Sees Positron Imaginary Advanced Field as Real Retarded Electron Field}

To understand the character of the advanced field of an electron, it is instructive to consider that to a time-reversed observer, the forward-time frame electron is a ``positron,'' and creates an imaginary retarded field. Further, the field the time-reversed observer computes is based on the positron having a negative, imaginary, fundamental (i.e. electron) charge.

So, forward and reversed time observers don't agree on the complexity of the charges.  What appears to the forward time observer as a real charge electron appears to the reverse time observer as an imaginary charge electron.  However the electron in the forward-time frame creates an imaginary retarded field in the time-reversed frame, which is viewed in the forward-time frame as an imaginary advanced field.  Thus, the souce of the advanced field is the same imaginary charge to both observers.

The imaginary or real character of the charge thus depends on which created field is under consideration.  A forward-time electron in the forward time frame is assigned a real character when creating a retarded field, but imaginary charge when creating an advanced field.  Conversely, the reversed-time electron in the forward time frame (i.e., a positron) creates a real advanced field and an imaginary retarded field.

At the same time, the real advanced field for an electron in the time forward frame is still a valid and relevant solution to the electromagnetic wave equation.  This field can be regarded (if desired) of causing the elctron motion that creates the retarded field.  Also, it is constructed by the real retarded field from the distant past absorber.

The imaginary constant on the field, coupled with the imaginary constant on the negative electron charge, leads to the positron apparent positive charge in the forward-time frame.  Also, the charge of a ``positron,'' a time-reversed electron, should be regarded by the forward-time observer as identical to that of an electron, in both sign and magnitude. The apparent positive charge of a time-reversed electron is due to the appearance of the imaginary constant in both the field itself (via the charge that created it) and with the charge it acts on.  An imaginary field acting on a real charge does not cause a real acceleration, nor does a real field acting on an imaginary charge.  But, it seems more parsimonius to consider charge an invariant to both observers, in magnitude and sign, while the choice in allocating an imaginary character determined by the time-sense of the observer and the particle carrying the charge.  

\subsection{Summary of rules of interaction of time-symmetric field and particle descriptions}

Observers may be considered to have opposite temporal directions. Then,

1.  A time-forward (i.e., Lorentz factor \(\ge 1\)) particle is influenced only by a retarded field

2.  A time-reversed (i.e., Lorentz factor \(\le -1\))  particle is influenced only by an advanced field

3.  Observers must regard advanced fields in their frame as imaginary

4.  Observers see all charges as the same sign (negative per convention) but particles with negative Lorentz factors have imaginary charge

5. Moving charges regardless of Lorentz factor sign create both retarded and advanced fields

\section{Complex Form of Modulation Factor Accounting for Absorption}

So far, the modulation function that has been shown to solve the free-particle Schroedinger equation is fundamentally a real quantity.  Just as real electromagnetic waves may be usefully considered as the real part of a complex function, the time-symmetric modulation function found without assuming Osiak relativity could be represented as a part of a complex function.  However, unlike the Schroedinger wavefunction, it is not fundamentally complex.  Therefore it is of interest to investigate if the imaginary character of time-advanced electromagnetic fields predicted by Osiak relativity leads to a complex modulation function that can be argued represents the Schroedinger wavefunction.

\subsection{Real Part of Complex Modulation Factor}

The modulation function found above, that solves the time-dependent free-particle Schroedinger equation, is a real function that is a sum of two real sinusoids, one based on a retarded interaction and the other based on an advanced interaction. Properly incorporating Osiak relativity into the description however requires that the advanced interaction acquires an imaginary constant factor.  In order to obtain the real part of a complex wavefunction that is similar to the real modulation function found above, it is necessary to have a second interaction with a real field that has a phase similarity to the direct advanced source field. Such a field component may exist if the past absorber radiated a retarded field, that arrives at source at the present with phase similarity to the source advanced field.  Such an interaction is arguably required from a requirement of symmetry of electrodynamics. If the retarded field may interact with an absorber in the distant future, then a time-symmetric electrodynamics arguably requires that the advanced field from the source may interact with the absorber in the distant past.

A time reversed observer sees absorption of the retarded (from the time-reversed perspective) field from the source and its re-emission by the absorber in their distant future. Assuming here complete absorption, the advanced fields from the absorber have a phase delay of \(\pi/2\) radians to the time reversed observer.  To the forward time observer, the reversed-time advanced field is the distant past absorber real retarded field coherently causing the source motion. The absorber is thus a coherent radiator that in the distant past radiated a field that may be regarded as causing the motion of the source in the present. That is, the retarded field of the absorber at the source, where the retarded and advanced path lengths are equal and cancel, is \(\pi/2\) radians advanced by the time-reversed absorption process, and so is in phase with the source acceleration.   This process may be regarded as both retarded and advanced fields being created by the source that is due to source motion, with both the retarded and advanced fields in addition being real.  The real advanced field is phase-identical with the imaginary advanced field, but carries energy toward the future rather than toward the past for the imaginary advanced field. 

Suppose the distance from the source to an absorber particle is \(R_a\).  An advanced wave from the source ``emitted'' at time \(t\) reaches the absorber particle at time \(t_a = t - R_a/c\).  The re-radiated retarded wave arrives at the source at time \(t = t_a + R_a/c\).  It arrives at a point along the ray between the source and absorber particle separated a distance \(R\) from the source at time \(t = t_a + (R_a - R)/c =  t - R_a/c + (R_a - R)/c  =  t  - R/c \).  The re-radiated retarded field is thus in phase with the retarded field from the source, in the region between the absorber charge and the source.

In order to write the field due to the distant-past absorber particle retarded field at an arbitrary point between the source and absorber particles, it is thus necessary to consider the round trip path difference, and its rate of change with position.  So, if \(R\) is the distance from the source particle to the field point between the source and absorber particles, then the time at which the field-exciting motion causing radiation along the path relative to the current time at the point, \(t\), is \(t = t + R/c \equiv t_{\textnormal{{\scriptsize adv}}}\). The retarded field from the absorber is thus phase identical to the advanced wave from the source, apart from the phase change associated with interaction with the absorber. 

It is well established, as described by Wheeler and Feynman, that complete absorbtion of the retarded field leads to net phase delay of the re-radiated field of \(\pi/2\) radians.  As noted above, symmetry considerations lead to an expectation that the change from the imaginary advanced wave to the real re-radiated retarded wave is an advancement under complete absorption of \(\pi/2\) radians.  With \(\sin( \theta + \pi/2) = \cos \theta  \) and \(  \cos (\theta + \pi/2) = -\sin \theta\),  the modulation factor for test zitter particle moving in the magnetic field of the past absorber retarded re-radiated field caused by the source advanced field may be written as

\begin{eqnarray}
 \mbox{\boldmath$\beta$}_t \cdot   \mbox{\boldmath$\beta$}_{s,{\textnormal{{\scriptsize pabs,ret}}}} \nonumber \\ =   [-\sin  \gamma \omega_z ( t - \mbox{\boldmath$\upsilon$} \cdot \mbox{\boldmath$r$}/c^2) ,\cos \gamma \omega_z ( t - \mbox{\boldmath$\upsilon$} \cdot \mbox{\boldmath$r$}/c^2) ,0] \nonumber \\ \cdot [-\sin(\omega_z t_{\textnormal{{\scriptsize adv}}} + \pi/2 + \phi_0),\cos(\omega_z t_{\textnormal{{\scriptsize adv}}} + \pi/2 + \phi_0),0]   \nonumber \\ =  [-\sin\gamma \omega_z ( t - \mbox{\boldmath$\upsilon$} \cdot \mbox{\boldmath$r$}/c^2) ,\cos \gamma \omega_z ( t - \mbox{\boldmath$\upsilon$} \cdot \mbox{\boldmath$r$}/c^2) ,0] \nonumber \\ \cdot [-\cos(\omega_z t_{\textnormal{{\scriptsize adv}}} + \phi_0),-\sin(\omega_z t_{\textnormal{{\scriptsize adv}}}+ \phi_0),0]  \nonumber \\  \equiv  -[-\sin \alpha,\cos \alpha ,0] \cdot [\cos\beta,\sin\beta,0] \nonumber \\ = \sin \alpha\cos\beta - \cos \alpha \sin\beta \nonumber \\ = \sin (\alpha - \beta)
\label{pabs_ret} 
\end{eqnarray}

with \(\alpha \equiv  \gamma \omega_z ( t - \mbox{\boldmath$\upsilon$} \cdot \mbox{\boldmath$r$}/c^2) \), and \(\beta \equiv  \omega_z t_{\textnormal{{\scriptsize adv}}} + \phi_0\).   Thus,

\begin{eqnarray}
 \mbox{\boldmath$\beta$}_t \cdot   \mbox{\boldmath$\beta$}_{s,{\textnormal{{\scriptsize pabs,ret}}}} = \nonumber \\ \sin \left[\gamma \omega_z( t - \mbox{\boldmath$\upsilon$} \cdot \mbox{\boldmath$r$}/c^2) - \omega_z t_{\textnormal{{\scriptsize adv}}} - \phi_0\right] \nonumber \\ 
=  \sin \left[\gamma \omega_z( t - \mbox{\boldmath$\upsilon$} \cdot \mbox{\boldmath$r$}/c^2) - \omega_z (t + r/c) t_{\textnormal{{\scriptsize adv}}} - \phi_0\right] \nonumber \\ 
=  \sin \left(\omega_z \left[(\gamma -1) t - \gamma\mbox{\boldmath$\upsilon$} \cdot \mbox{\boldmath$r$}/c^2 -  r/c)\right]  - \phi_0\right).
\label{retarded_modfact_tabsret}
\end{eqnarray}

The real part of the modulating factor on the radial inverse-square magnetic force for the test zitter particle moving in the sum of the retarded direct and advanced source to retarded past absorber fields is then, with \(  \cos (\theta - \pi/2) = \sin \theta\)

\begin{eqnarray}
{\cal M}_{{\textnormal{{\scriptsize sym,real}}}}(t) \equiv  \mbox{\boldmath$\beta$}_t \cdot   \mbox{\boldmath$\beta$}_{s,{\textnormal{{\scriptsize ret}}}}  + \mbox{\boldmath$\beta$}_t \cdot \mbox{\boldmath$\beta$}_{s,{\textnormal{{\scriptsize pabs,ret}}}}   \nonumber \\ 
=  {\cal M}_{{\textnormal{{\scriptsize ret}}}}(t)  + {\cal M}_{{\textnormal{{\scriptsize pabs,ret}}}}(t)   = {\cal M}_{{\textnormal{{\scriptsize sym,real}}}}(\mbox{\boldmath$r$},t) \nonumber \\ 
= \cos ( \omega_z \left[ (\gamma - 1) t - \gamma  \mbox{\boldmath$\upsilon$} \cdot \mbox{\boldmath$r$}/c^2 +  r/c\right]  - \phi_0) \nonumber \\
 + \sin ( \omega_z \left[ (\gamma - 1) t - \gamma  \mbox{\boldmath$\upsilon$} \cdot \mbox{\boldmath$r$}/c^2 -  r/c\right]  - \phi_0)  \nonumber \\
= \cos ( \omega_z \left[ (\gamma - 1) t - \gamma  \mbox{\boldmath$\upsilon$} \cdot \mbox{\boldmath$r$}/c^2 +  r/c\right]  - \phi_0)\nonumber \\ + \cos ( \omega_z \left[ (\gamma - 1) t - \gamma  \mbox{\boldmath$\upsilon$} \cdot \mbox{\boldmath$r$}/c^2 -  r/c\right]  - (\pi/2  + \phi_0)) 
\end{eqnarray}

 Using the identity \(\cos\theta + \cos\phi = 2\cos((\theta + \phi)/2)\cos((\theta - \phi)/2) \):
\begin{eqnarray}
{\cal M}_{{\textnormal{{\scriptsize sym,real}}}}(t) = \nonumber \\ 
2\cos ( \omega_z \left[ (\gamma - 1) t - \gamma  \mbox{\boldmath$\upsilon$} \cdot \mbox{\boldmath$r$}/c^2\right]  - \pi/4  -\phi_0 )\nonumber \\ \cdot \cos ( \omega_z  r/c  + \pi/4 ) 
\end{eqnarray}

Comparison with Eq. (\ref{SSTimeSymmetricFactor}) finds that the previous results obtained assuming real-definite electromagnetic fields of Eqs. (\ref{TDSlikeEq_osiak}) and (\ref{TIndSlikeEq_osiak}) for the real part of the free-particle time-dependent and time-independent Schroedinger equations are recovered.

\subsection{Imaginary Part of Complex Modulation Factor}

Due to complete absorption of the retarded field from the source and its re-emission by the absorber in the distant future, the fields from the absorber have a phase delay of \(\pi/2\) radians.  That is, the advanced field of the absorber at the source, where the retarded and advanced path lengths are equal and cancel, is \(\pi/2\) radians behind the source motion.   This process is identical to both retarded and advanced fields being created by the source that is due to source motion delayed by \(\pi/2\) radians from its actual motion, with both the retarded and advanced fields in addition being imaginary.

The imaginary advanced field of the future absorber carrying energy toward the source is phase-delayed by \(\pi/2\) radians compared to a re-radiated field by an absorber particle on its inner boundary.  With the phase delay  averaged over the absorber of \(\pi/2\) radians, it is phase-identical to the retarded field from the source. This is the field that in WFAT reinforces the half-retarded field of the source, in the region between the source and the absorber particle.  According to Osiak relativity it has a separate identity due to its imaginary character, and it is transporting energy toward the source rather than away.   The modulation due to this field part is, with \(  \cos (\theta - \pi/2) = \sin \theta\), and assuming the source zitter particle is stationary so that \(r_{\textnormal{{\scriptsize ret}}}(t) = r(t) \equiv r\),
\begin{eqnarray}
 \mbox{\boldmath$\beta$}_t \cdot   \mbox{\boldmath$\beta$}_{s,{\textnormal{{\scriptsize fabs,adv,in}}}} =   \nonumber \\ 
\cos \left[\gamma \omega_z( t - \mbox{\boldmath$\upsilon$} \cdot \mbox{\boldmath$r$}/c^2) - \omega_z t_{\textnormal{{\scriptsize ret}}} - \pi/2 - \phi_0\right] \nonumber \\
=   \sin \left[\gamma \omega_z( t - \mbox{\boldmath$\upsilon$} \cdot \mbox{\boldmath$r$}/c^2) - \omega_z t_{\textnormal{{\scriptsize ret}}} - \phi_0\right]\nonumber \\
=  \sin ( \omega_z \left[ (\gamma - 1) t - \gamma  \mbox{\boldmath$\upsilon$} \cdot \mbox{\boldmath$r$}/c^2 +  r/c\right] - \phi_0), 
\label{retarded_modfact_tabsret}
\end{eqnarray}
where the subscript notation ``s'' is for ``source,'' ``fabs'' is for ``future absorber,'' and ``in'' is for ``inbound'' to the source particle from the point of view of the reversed-time observer.  The designation ``inbound'' indicates that at a particular field point, the field component here is due to an absorber paricle that is beyond the field point from the source position.  (``Outbound'' from the reverse-time perspective will refer to the inverted field after passing through the source, as discussed below.) Although this field component is an advanced field and so imaginary, it is in-phase to the source retarded field but delayed by \(\pi/2\) radians.

It can be noted that in WFAT, the ``inbound'' advanced field from the future absorber reinforces the source half-retarded field to full strength.  Such reinforcement does not occur here in the present model on account of the advanced field being imaginary.  

The modulating factor for the spin-aligned test zitter particle moving in the inbound future absorber field is thus
\begin{eqnarray}
{\cal M}_{{\textnormal{{\scriptsize fabs,adv,in}}}}(t) = \nonumber \\ \sin ( \omega_z \left[ (\gamma - 1) t - \gamma  \mbox{\boldmath$\upsilon$} \cdot \mbox{\boldmath$r$}/c^2 +  r/c\right] - \phi_0) . 
\label{retarded_modfact_tabsret_2} 
\end{eqnarray}

Had previously (\ref{modfact_advanced}) (for stationary source zitter particle), for the test zitter particle moving in the direct advanced field,

\begin{equation}
{\cal M}_{{\textnormal{{\scriptsize s,adv}}}}(t) = \cos ( \omega_z \left[ (\gamma - 1) t - \gamma  \mbox{\boldmath$\upsilon$} \cdot \mbox{\boldmath$r$}/c^2 -  r/c\right]  - \phi_0).
\label{modfact_advanced_copy} 
\end{equation}

This can be rewritten, with \(\sin \theta = \cos(\theta - \pi/2)\), as

\begin{equation}
{\cal M}_{{\textnormal{{\scriptsize s,adv}}}}(t) = \sin ( \omega_z \left[ (\gamma - 1) t - \gamma  \mbox{\boldmath$\upsilon$} \cdot \mbox{\boldmath$r$}/c^2 -  r/c\right] - \pi/2   - \phi_0).
\label{modfact_advanced_copy} 
\end{equation}

So, with \(\sin\theta + \sin\phi = 2\sin((\theta + \phi)/2)\cos((\theta - \phi)/2) \),
\begin{eqnarray}
 \mbox{\boldmath$\beta$}_t \cdot \mbox{\boldmath$\beta$}_{s,{\textnormal{{\scriptsize fabs,adv,in}}}} + \mbox{\boldmath$\beta$}_t \cdot   \mbox{\boldmath$\beta$}_{s,{\textnormal{{\scriptsize adv}}}} = \nonumber \\ {\cal M}_{{\textnormal{{\scriptsize fabs,adv,in}}}}(t)  +  {\cal M}_{{\textnormal{{\scriptsize adv}}}}(t) \nonumber \\ \equiv {\cal M}_{{\textnormal{{\scriptsize sym,imag}}}}(t) = {\cal M}_{{\textnormal{{\scriptsize sym,imag}}}}(\mbox{\boldmath$r$},t) \nonumber \\ 
= \sin ( \omega_z \left[ (\gamma - 1) t - \gamma  \mbox{\boldmath$\upsilon$} \cdot \mbox{\boldmath$r$}/c^2 +  r/c\right] - \phi_0) \nonumber \\ 
+  \sin ( \omega_z \left[ (\gamma - 1) t - \gamma  \mbox{\boldmath$\upsilon$} \cdot \mbox{\boldmath$r$}/c^2 -  r/c\right]  - \phi_0 - \pi/2) \nonumber \\ 
= 2 \sin \left(\omega_z \left[ (\gamma - 1) t - \gamma  \mbox{\boldmath$\upsilon$} \cdot \mbox{\boldmath$r$}/c^2\right]   - \pi/4 - \phi_0\right) \nonumber \\ \times \cos \left(\omega_z r/c  + \pi/4\right). 
\label{SSTimeSymmetricFactorCopy}
\end{eqnarray}

\subsection{Complex Modulation Factor for Aligned Spins}

The complex modulation that is comparable to the complex Schroedinger wavefunction may now be constructed from the real and imaginary modulation factors as  

\begin{widetext}

\begin{eqnarray}
{\cal M}_{{\textnormal{{\scriptsize cmplx}}}}(\mbox{\boldmath$r$},t)   \equiv {\cal M}_{{\textnormal{{\scriptsize sym, real}}}}(\mbox{\boldmath$r$},t)   + i{\cal M}_{{\textnormal{{\scriptsize sym,imag}}}}(\mbox{\boldmath$r$},t)  \nonumber \\ 
 = 2\cos ( \omega_z \left[ (\gamma - 1) t - \gamma  \mbox{\boldmath$\upsilon$} \cdot \mbox{\boldmath$r$}/c^2\right]  - \pi/4 -\phi_0 )  \times \cos ( \omega_z  r/c + \pi/4 ) 
\nonumber \\ 
+ i 2 \sin \left(\omega_z \left[ (\gamma - 1) t - \gamma  \mbox{\boldmath$\upsilon$} \cdot \mbox{\boldmath$r$}/c^2\right]  - \pi/4 - \phi_0\right)  \times \cos \left(\omega_z r/c  + \pi/4\right)
\nonumber \\ 
= 2\exp i( \omega_z \left[ (\gamma - 1) t - \gamma  \mbox{\boldmath$\upsilon$} \cdot \mbox{\boldmath$r$}/c^2\right]  - \pi/4  -\phi_0 )  \times \cos ( \omega_z  r/c  + \pi/4 ) 
\nonumber \\ 
 = 2\exp i \omega_z \left[ (\gamma - 1) t - \gamma  \mbox{\boldmath$\upsilon$} \cdot \mbox{\boldmath$r$}/c^2\right]  \times \exp -i(\pi/4  + \phi_0 ) \cos ( \omega_z  r/c  + \pi/4 ) 
\label{complex_sym_modfact} 
\end{eqnarray}
\end{widetext}
Performing variable substitutions as shown above in section \ref{SepVarOsiak}  , the first factor on the right hand side of the final equation of (\ref{complex_sym_modfact}) can be shown to solve the free-particle time-dependent Schroedinger equation. The other factors separate out.

\section{Interpretation}
\label{sectdisc}

The central result of the work presented herein may be stated as follows. If classical Hamiltonian mechanics is valid, the position and momentum dependent modulation factor on the time-symmetric Coulomb-like magnetic force between classical Dirac particles satisfies a time-independent auxiliary equation that may be solved to find constraints on allowed particle trajectories. Assuming the present standard formulation of relativity theory, the time-independent auxiliary equation is formally similar, but not identical, to the time-independent Schr\"odinger equation. Using an alternative relativity theory formulation suggested by Osiak,  the resulting auxiliary equation is formally identical to the time-independent Schr\"odinger equation. 

The Schr\"odinger wavefunction can thus be regarded as a kinematical feature of classical physics and Hamiltonian mechanics, assuming time-symmetric electromagnetic interaction and a real zitterbewegung. The wavefunction is not a dynamical field quantity that influences particle motions. Rather, it is an indication that classical interactions between particles with intrinsic spin are more complicated and interesting than has been previously realized.

\section{Concluding Remarks}
\label{sectconc}

A plausible classical physics based explanation for quantum behavior is straightforwardly obtainable assuming time-symmetric electromagnetic interactions of charged point particles undergoing relativistic circulatory motions accounting for intrinsic spin.  This explanation is so far incomplete, but deserves further consideration and investigation due to its explanatory power and more parsimonious nature than the currently-established quantum theory. 

No new hypotheses beyond classical electrodynamics are required to reach the results provided. The existence of the electron intrinsic spin and magnetic moment are taken to be empirical fact and dynamical behaviors, not fundamental attributes. The hypothesis that time-advanced interaction terms arising naturally in the classical description are to be disregarded is not used.  


\end{document}